\documentclass[11pt]{article}

\usepackage[iso-8859-7]{inputenc}
\usepackage{epsfig}
\usepackage{graphicx}
\usepackage{epstopdf}
\usepackage{amsfonts}
\usepackage{amssymb}
\usepackage{amsthm}
\usepackage{amsmath}
\usepackage{multirow}

\usepackage{color}

\newcommand{\beq}{\begin{equation}}
\newcommand{\eeq}{\end{equation}}
\newcommand{\bea}{\begin{eqnarray}}
\newcommand{\eea}{\end{eqnarray}}

\newcommand{\gsim}{\lower.7ex\hbox{$\;\stackrel{\textstyle>}{\sim}\;$}}
\newcommand{\lsim}{\lower.7ex\hbox{$\;\stackrel{\textstyle<}{\sim}\;$}}
\newcommand{\nnmb}{\nonumber}

\newcommand{\be}{\begin{equation}}
\newcommand{\ee}{\end{equation}}
\newcommand{\ba}{\begin{eqnarray}}
\newcommand{\ea}{\end{eqnarray}}

\usepackage{float}
\usepackage{amsmath} 
\usepackage{amssymb}   
\usepackage{amsthm} 
\usepackage{geometry}
\usepackage{graphicx} 
\usepackage{multirow} 
\begin{document} 
	\thispagestyle{empty}

	\begin{titlepage}
		
		%\vspace*{-15mm}
		%\begin{flushright}
		%SHEP-11-XX\\
		%\end{flushright}
		\vspace*{0.7cm}

		\begin{center}
			{\Large {\bf Perturbative moduli stabilisation in type IIB/F-theory framework}}
			\\[12mm]
		Ignatios Antoniadis$^{a,b}$~\footnote{E-mail: \texttt{antoniad@lpthe.jussieu.fr}}, 
			Yifan Chen$^{a}$~\footnote{E-mail: \texttt{yifan.chen@lpthe.jussieu.fr}},
			George K. Leontaris$^{c}$~\footnote{E-mail: \texttt{leonta@uoi.gr}}
			\\[-2mm]
					\end{center}
		\vspace*{0.50cm}
	\centerline{$^{a}$ \it
Laboratoire de Physique Th\'eorique et Hautes \'Energies - LPTHE,
}
		\centerline{\it
			Sorbonne Universit\'e, CNRS, 4 Place Jussieu, 75005 Paris, France}
\vspace*{0.2cm}
\centerline{$^{b}$ \it
Albert Einstein Center, Institute for Theoretical Physics, University of Bern,}
\centerline{\it
Sidlerstrasse 5, CH-3012 Bern, Switzerland}
\vspace*{0.2cm}
\centerline{$^{c}$ \it
Physics Department, University of Ioannina}
\centerline{\it 45110, Ioannina, 	Greece}
\vspace*{1.20cm}

\begin{abstract}
\noindent

We propose a new mechanism of (geometric) moduli stabilisation in type IIB/F-theory four-dimensional compactifications on Calabi-Yau manifolds, in the presence of $7$-branes, that does not rely on non-perturbative effects. Complex structure moduli and the axion-dilaton system are stabilised in the standard way, without breaking supersymmetry, using 3-form internal fluxes. K\"ahler class moduli stabilisation utilises perturbative string loop corrections, together with internal magnetic fields along the $D7$-branes world-volume leading to Fayet-Iliopoulos D-terms in the effective supergravity action. The main ingredient that makes the stabilisation possible at a de Sitter vacuum is the logarithmic dependence of the string loop corrections in the large two-dimensional transverse volume limit of the $7$-branes.

\end{abstract}
\end{titlepage}

\section{Introduction}
The String Theory landscape comprises an enormous number of vacua, however, not all 
of them are consistent with the  cosmological data and the relevant for particle physics effective  $N=1$ supergravity theories.
Many of them are characterised by anti-de-Sitter (AdS) minima, predicting a negative cosmological 
constant, in contradiction with the existing evidence of the accelerated expansion of the universe. 
In order to obtain a  consistent  supersymmetric vacuum we must seek string compactifications
with a  de Sitter (dS) minimum and stabilise the various moduli fields which are ubiquitous in
string compactifications. 

In type IIB string theory in particular, compactified on a Calabi-Yau three-fold, the complex structure moduli and the axion-dilaton/ten-dimensional (complexified) string coupling appear in the superpotential induced when 3-form fluxes are turned on and can be fixed in a supersymmetric way~\cite{Frey:2002hf} \footnote{For an alternative stabilisation method, see \cite{Dasgupta:1999ss, Taylor:1999ii}}. The K\"ahler class moduli on the other hand, such as the Calabi-Yau volume, remain undetermined because -being $(1,1)$ forms- they do not appear in the flux induced superpotential. The resulting effective supergravity has constant superpotential and, thus, a non-vanishing gravitino mass term and vanishing scalar potential, due to the no-scale structure of the tree-level K\"ahler potential of the K\"ahler class moduli that remain massless and undetermined.
Their stabilisation requires their appearance in the superpotential
and, a usual way to realise it, is to include non-perturbative corrections~\cite{Kachru:2003aw}. 
These are in general model dependent related, for instance, to gaugino condensation of the gauge group 
of $D7$-branes~\cite{Derendinger:1985kk}.  In the simplest case, to realise a sufficiently large volume in a well controlled regime, a fine tuning of the coefficients in the resulting superpotential generated by the fluxes is required.

Moreover, higher order $\alpha '$ corrections are taken into account and 
break the no-scale structure of the K\"ahler potential~\cite{Becker:2002nn}.  
One-loop corrections to the K\"ahler potential may also be included ~\cite{vonGersdorff:2005bf, Berg:2005ja, Berg:2005yu, Parameswaran:2006jh, Cicoli:2007xp, Berg:2014ama, Haack:2015pbv, Kobayashi:2017zfd, Haack:2018ufg}. In most cases,  
dS vacua can only be obtained by `uplifting' the vacuum energy in the presence of anti-$D3$ branes ($\overline{D3}$-branes for short), which  break though supersymmetry explicitly (KKLT scenario~\cite{Kachru:2003aw}). This situation can, in principle, be remedied
if instead  of $\overline{D3}$-branes, 
D-term contributions are taken into account in the effective action~\cite{Burgess:2003ic, Jockers:2004yj, Antoniadis:2004pp, Antoniadis:2005nu, Antoniadis:2008uk, Achucarro:2006zf, Haack:2006cy}, emerging from internal magnetic fluxes along the $D7$-branes world volume. String realisations improving  the KKLT scenario are also possible within the so-called large volume scenario in
Calabi-Yau compactifications~\cite{Balasubramanian:2005zx, Conlon:2005ki}.

In the present work we take a different path and, working in the framework of type IIB/F-theory,
we consider possible contributions to the K\"ahler potential due to the presence of space-time filling  $D7$-branes.  
Recall that 7-branes are fundamental objects in F-theory and certain configurations of them 
determine the gauge symmetry of the effective theory. Moreover, an important class of  matter fields 
resides on Riemann surfaces  which can be interpreted  as the intersection locus of $D7$-branes.
In such configurations, anomalous $U(1)$ symmetries, associated with intersecting branes,
are frequently present and the resulting chiral spectrum of the four-dimensional theory usually induces Fayet-Iliopoulos (FI) D-terms to the effective potential. 
The importance of introducing  D-term  contributions  is that these are always positive and can in principle uplift  the potential,  generating a dS minimum with all moduli fixed. 

It turns out though that D-terms are not sufficient to stabilise all moduli, at least when charged fields on $D7$-branes have vanishing expectation values. The basic additional ingredient that comes in rescue is quantum corrections to the K\"ahler potential stemming from the presence of the $D7$-branes. For large transverse volume, as is the case of such configurations, these corrections become crucial and cannot be neglected. 
They display a logarithmic dependence on the modulus  associated  with the transverse dimension~\cite{Antoniadis:1998ax}. The logarithmic dependence is quite general in the presence of two large transverse dimensions, as shown for example in~\cite{Antoniadis:2002gw}. Combining these effects with  D-term contributions, we show that all K\"ahler moduli can be stabilised in a dS vacuum of broken supersymmetry.

The layout of this paper is as follows. In Section 2, we start with a general overview of the moduli stabilisation problem, we introduce the effective supergravity and present the leading quantum corrections, perturbative in $\alpha'$ and in the string coupling (subsection 2.1). We discuss in particular their dependence on the transverse volume of the $D7$-branes that grows logarithmically at large distances. We then introduce the D-term contributions to the effective potential (subsection 2.2).  In Section 3, we work out the minimisation conditions and the K\"ahler moduli stabilisation. We first present the simple example of a single $D7$-brane which brings two moduli that can be chosen to be the total volume of the Calabi-Yau manifold and the volume transverse to the brane (subsection 3.1). We show how the latter can be stabilised using the logarithmic loop corrections, but not the former. Full stabilisation in a dS minimum can be achieved only in the general case of three intersecting $D7$-branes with corresponding D-terms. Indeed, the total volume can be stabilised by the logarithmic corrections (subsection 3.2), while all moduli are fixed when D-terms are included, as shown in subsection 3.3. 
Finally, Section 5 contains our conclusions, while in the Appendix we show why in the case of one $D7$-brane, one cannot find a dS minimum in the whole parameter space of the model.

\section{General  Overview}

In IIB string theory, appropriate 3-form fluxes generate a superpotential given by~\cite{Gukov:1999ya}:
\be   {\cal W} = \int G_3\wedge \Omega~\cdot \label{supII}
\ee
In the above, the $G_3$ flux is defined as $G_3= F_3- SH_3$, where $F_3, H_3$ are the Ramond-Ramond (RR) and Neveu-Schwarz (NS) 3-form fluxes, 
$S=C_0+i e^{-\phi}\equiv C_0+i/g_s$ is the axion-dilaton field associated with the ten-dimensional string coupling $g_s$,  and $\Omega$ is the holomorphic Calabi-Yau $(3,0)$-form dependent on complex structure moduli $z_a$ \cite{Candelas:1990pi}. 
Clearly, the superpotential (\ref{supII}) 
depends on $z_a$ and the axion-dilaton $S$ but it  is independent of the K\"ahler class moduli, as described for example in~\cite{Giddings:2001yu}.
The conditions $D_a{\cal W}=0$, (where $D_a$ is the K\"ahler covariant derivative and the index $a$ runs over all moduli fields)
fix all the complex structure moduli and the axion-dilaton.  A generalization of eq.~(\ref{supII}) in the F-theory framework is straightforward~{\cite{Lust:2005bd, Honma:2017uzn}.

A wide class of solutions towards the stabilisation of K\"ahler class moduli rely on non-perturbative effects. This ingredient allows the appearance of the K\"ahler moduli in the superpotential and as a 
result, a potential is generated whilst their masses are determined  from the minimisation procedure of the effective potential. The $\alpha'$-corrections which 
generate ${\cal O}({\alpha'}^3)$ contributions to the K\"ahler potential, are also widely used. In the present 
analysis, we will take a different path and investigate the effects of string loop corrections to the K\"ahler potential which displays a dependence on the transverse 
volume of $D7$-branes,  as well as the D-terms potential depending  on the world-volume of  $D7$-branes.

To set the stage, we start with the K\"ahler potential. Ignoring for the moment $\alpha'$ or string loop corrections, it can be written as a simple separable form
\be 
{\cal K}_0 = {\cal K}_0 (T_i) + {\cal K}(S, z_a)~, \label{K0}
\ee 
where $T_i$ are the K\"ahler moduli, $S$  the axion-dilaton $S=C_0+ie^{-\phi}$, and   $z_a$  the complex structure moduli respectively. The two components of ${\cal K}$  are
\ba
{\cal K}_0 (T_i)&=& -\sum_{i = 1}^3 \ln(-i(T_i-\bar{T_i})) \label{noscale}\\
{\cal K}(S, z_a)&=& - \ln(-i(S-\bar{S}))-\ln(i\int\Omega\wedge \bar{\Omega})~\cdot
\ea
%\noindent
As can be readily seen, eq. (\ref{noscale}) satisfies the no-scale condition
\beq 
\sum_{I, J = T_i} \mathcal{K}_0^{I \overline{J}} \partial_I \mathcal{K}_0 \partial_{\overline{J}}\mathcal{K}_0 = 3,
\eeq
and therefore  the potential can be written as
\be
V= \sum_{I, J \neq T_i}e^{\cal K}\left(D_I \mathcal{W} {\cal K}_{I\bar J}^{-1} D_{\bar J} \mathcal{W}\right),
\ee
with $D_I \mathcal{W}=\partial_I \mathcal{W} + \mathcal{W} \partial_I{\cal K}$.
In the simplest scenario, the flux generated perturbative superpotential in eq. (\ref{supII}) stabilises the
complex structure moduli and the axion-dilaton field by using the supersymmetric conditions $D_i \mathcal{W} = 0$. This leads to zero vacuum energy. Furthermore, 
because of the no scale structure of the K\"ahler potential, the K\"ahler moduli cannot be fixed from the 
supersymmetric minimisation of the superpotential.
In order to stabilise the K\"alher moduli, quantum corrections are usually included so that the no-scale invariance is violated and the total volume is fixed. 

 In addition, non-perturbative contributions
originating from gaugino condensation or instanton effects, involve certain  K\"ahler moduli 
in exponentially suppressed terms.  Including such terms, the superpotential obtains the form
\be
{\cal W} = \mathcal{W}_0+\sum_{i = 1}^{h_+^{1, 1}}\Lambda_i e^{ - \lambda_i T_i}~\cdot \label{W0AT}
\ee
The implementation of the conditions $D_{z_a}{\cal W}=0$  put all complex structure  moduli and axion-dilaton at their minima, 
and so $\mathcal{W}_0$, $\Lambda_i$ and $\lambda_i$ are constants.
The condition $D_{T_i}{\cal W}=0$ leads to a supersymmetric AdS minimum~\cite{Kachru:2003aw}.
However, a drawback of this scenario is that the minimum is achieved only when the value of $\mathcal{W}_0$ is 
fine-tuned in order to balance the non-perturbative effects. In addition, 
as can be seen from (\ref{W0AT}), non-perturbative corrections are required for all 4-cylces 
involved, whilst the inclusion of $\overline{D3}$ contributions to uplift the AdS minimum
breaks supersymmetry explicitly.

A generalisation of the above scenario \cite{Balasubramanian:2005zx, Conlon:2005ki} improving these deficiencies,
is realised with an exponentially large volume, where in the simplest 
case of two K\"ahler moduli $\tau_b, \tau_s$ the Calabi-Yau volume takes the form ${\cal V}= \tau_b^{3/2}-\tau_s^{3/2}$. 
In the presence of  non-perturbative corrections the K\"ahler potential
and superpotential are given by
\bea
 \mathcal{K}_{LVS} &&=  - 2 \textrm{ln} (\tau_b^{\frac{3}{2}} - \tau_s^{\frac{3}{2}} + \xi),\\
\mathcal{W}_{LVS} &&= \mathcal{W}_0 + \Lambda e^{- \lambda\tau_s}.\label{LVS}
\eea
The gaugino condensation and the $\alpha^\prime$ correction $\xi$ are necessary to stabilise both $\tau_b$ and $\tau_s$. However, as in the  KKLT case, a mechanism is required to uplift it to a dS  minimum.

In the next subsections, we will present an alternative scenario of K\"ahler moduli stabilisation which does not rely on (uncontrolable)
non-perturbative corrections in the superpotential. The proposed mechanism is based on the observation that  
the effective action receives logarithmic corrections in the large (two dimensional) volume limit transverse to the $D7$-branes~\cite{Antoniadis:1998ax}.

\subsection{Effective supergravity, dualities and quantum corrections}

We would like now to include the leading quantum corrections in the effective action presented above. 
These are perturbative in $\alpha'$ or in the string coupling correcting the K\"ahler potential and the gauge
kinetic functions. The former correspond to a constant shift of the internal volume proportional to the Euler number of the Calabi-Yau manifold, while the latter can be important only in the presence of large transverse volume of dimension less or equal than two, as in the configuration 
of $D7$-branes that we consider in this work. In this case, one loop corrections to localised effective action terms in the open string channel grow logarithmically with the transverse volume~\cite{Antoniadis:1998ax}. Since such corrections have been computed explicitly in $N=1$ type I orientifolds~{\cite{Antoniadis:1996vw}, we will start by presenting them in this framework and then use T-dualities to derive the corresponding expressions in type IIB/F-theory context with intersecting $D7$-branes. 

Let us consider type I strings on a product of three 2-torii $(\prod_{i=1}^3 T^2_i)$ with $D9$-branes and in general three types of $D5$-branes extended in the three non-compact spatial dimensions and along each one of the three $T^2_i$. The K\"ahler potential is 
\ba 
{\cal K}=-\ln (S-\bar S)-\sum_{i=1}^{3}\ln(T_i-\bar T_i)+\cdots,
\label{KahlerTreeI}
\ea 
where the dots refer to contributions dependent on the complex structure that we omit in the following. The imaginary part of the various moduli are given by the inverse gauge couplings of the $D9$ and $D5_i$ branes, upon compactification in four dimensions:
\ba
{\rm Im}S={1\over g_9^2}=e^{-\phi}v_1v_2v_3\qquad {\rm Im}T_i={1\over g_{5i}^2}=e^{-\phi}v_i~,
\ea
with $v_i$ the volume of $T^2_i$ in string units. 

To go to the framework of type IIB/F-theory with three types of $D7$-branes, one has to perform six T-dualities along all six internal directions. The $D9$ then becomes $D3$ while $D5_i$ becomes $D7_i$ transverse to $T^2_i$.
Recall that under a single T-duality $R\to 1/R$ the string coupling transforms as $e^\phi\to e^\phi/R$. It follows that under six T-dualities, the four moduli $S,T_i$ go to the inverse gauge couplings of the corresponding $D3$ and $D7_i$ branes:
\ba
{\rm Im}S\to {1\over g_3^2}=e^{-\phi}\qquad {\rm Im}T_i\to {1\over g_{7i}^2}=e^{-\phi}{\cal V}/v_i
\ea
with ${\cal V}=v_1v_2v_3$ the total internal volume and $e^\phi$ is the 10-dimensional string coupling $g_s=e^\phi$. The K\"ahler potential (\ref{KahlerTreeI}) then becomes
\ba 
{\cal K}\to -2\ln (e^{-2\phi} {\cal V}) = -\ln (S-\bar S) -2\ln \hat{\cal V}
\label{KahlerTreeII}
\ea 
where $\hat{\cal V}=e^{-3\phi/2}{\cal V}$.

Corrections to the K\"ahler potential  in type II strings (with D-branes), are induced through corrections of the Einstein graviton kinetic terms. The perturbative corrections in $\alpha'$, $\hat\xi$, and the string one-loop corrections $\hat\delta$, both arise  in the string frame as corrections to the Einstein kinetic terms~\cite{Antoniadis:2002tr}:
\ba\label{hatcorrections}
\left[ e^{-2\phi}({\cal V}+\hat\xi) + \hat\delta\right]{\cal R}\label{sfc}\,
\ea
where $\hat\xi$ is of order $\alpha'^3$, arising at four loops in the Calabi-Yau $\sigma$-model, and is proportional to the Euler number of the Calabi-Yau manifold $\chi$, $\hat\xi=-\left[\zeta(3)/4(2\pi)^3\right]\chi$~\cite{Grisaru:1986kw, Antoniadis:1997eg, Becker:2002nn}, and $\hat\delta$ is in general a function of moduli fields. 

From the above form, it follows that the two corrections can be accounted for by a shift of the Calabi-Yau volume $\cal V$ and of the inverse 4d closed string coupling $e^{-2\phi_4}=e^{-2\phi}{\cal V}$:
\ba
{\cal V}\to{\cal V} +\xi\qquad ;\qquad
e^{-2\phi_4}=e^{-2\phi}{\cal V}\to e^{-2\phi_4} +\hat\delta
\ea
It is now clear that the radiatively corrected K\"ahler potential reads:
\ba\label{Kahlercorrected}
{\cal K}&=&-2\ln\left[e^{-2\phi}({\cal V}+\hat\xi)+\hat\delta\right] \nonumber\\
&=&-\ln e^{-\phi}-2\ln\left(\hat{\cal V}+\xi+\delta\right)\nonumber\\
&=&-\ln(S-\bar S)-2\ln\left(\hat{\cal V}+\xi+\delta\right) \, ,
\ea
where $\hat{\cal V}$ is defined in (\ref{KahlerTreeII}) and
\ba\label{unhated}
\xi=\hat\xi/g_s^{3/2}= -\frac{\zeta(3)}{4(2\pi)^3g_s^{3/2}}\chi
\quad;\quad \delta=\hat\delta g_s^{1/2}\, .
\ea
Note that a nonvanishing $\xi$ in the large volume limit gives rise to localised graviton kinetic terms in the internal Calabi-Yau space at the points where the Euler number is concentrated. Indeed it remains finite in the large volume limit in (\ref{sfc}) leading to a localised Einstein action, on an effective $3$-brane, studied in \cite{Antoniadis:2002tr}.  

These localised graviton kinetic terms can generate one loop corrections that grow logarithmically with the size of the bulk in the presence of 7-brane sources~\cite{Antoniadis:1998ax}. Indeed, the localised graviton vertices can emit closed strings propagating along all the six dimensions of the internal space. The contribution of the relevant diagrams contains the exchange of these closed strings in the bulk between a certain number of graviton vertices from the 4d Einstein action localised in the internal space and another boundary that can be a $D$-brane or an orientifold plane. These diagrams correspond to local tadpoles whose existence can be consistent with global tadpole cancellation. Each of branes/orientifold planes behave as  point-like sources in the corresponding transverse space. The emitted closed string from the localised graviton vertices carry in principle momentum along all the six internal dimensions. However, the momentum along the directions parallel to the worldvolume of brane/orientifold plane vanishes by conservation. It follows that the exchanged closed strings carry only transverse momentum $p_\perp$ which is not conserved due to the presence of branes/orientifold planes that break translation invariance in the transverse directions.  Thus the relevant diagram that contributes to $\delta$, in the large transverse volume $V_\perp$ limit, takes the form: 
\be
\delta \sim \frac{1}{V_\perp}\sum_{|p_\perp|<M_s} \frac{1}{p_\perp^2} F(\vec{p}_\perp) \qquad;\qquad \vec{p}_\perp=\left(\frac{n_1}{R},\cdots,\frac{n_d}{R}\right)\, ,\label{TA}
\ee
where $F(\vec{p}_\perp)$ are the local tadpoles in the momentum space and the summation (instead of integration) is because $\vec{p}_\perp$ are discrete in the compact transverse space that we parameterise its size as $V_\perp \sim R^d$. The tadpoles arise from the distribution of D-branes and orientifolds which act as classical point-like sources in the transverse space. Considering for instance $2^d$ orientifolds located at the corners of a $d$-dimensional cube formed by $d$ dimensions of equal size $\pi R$ and a brane at the position $\vec{y}$ (plus its images), the local tadpole is given by:
\be
F(\vec{p}_\perp) \sim \left\{ \prod_{i=1}^d \left(\frac{1+(-)^{n_i}}{2}\right)-\cos(\vec{p}_\perp \vec{y})\right\}.
\ee
It follows that its contribution to the amplitude (\ref{TA}) would contain an infrared divergence in the large transverse limit when its co-dimension is less or equal to 2. The divergence is linear in $R$ for $d=1$ and logarithmic for $d=2$, while the amplitude is finite for $d>2$. 

In conclusion, in the  system with 7-branes and localised graviton kinetic terms in the internal space, the effective two-dimensional propagation of closed strings induce an infrared divergence in the loop correction that goes logarithmically when the co-dimension 2 transverse dimension is large \cite{Antoniadis:1998ax}. Due to the infrared divergence, one could also expect it is the dominant correction at that order in the string loop expansion. One could thus write eq. (\ref{TA}) as
\be
\delta = \eta\ln u, \label{logdelta}
\ee
where $\eta$ is some model dependent constant and $u$ is the modulus of the space transverse to a D7-brane.

We would like to emphasise again that the necessary condition for the arguments of ~\cite{Antoniadis:1998ax} is to have localised kinetic terms in the internal space. Here we have to discuss separately the case of smooth Calabi-Yau manifolds and orbifolds. As argued before, the presence of a non-vanishing $\xi$ at the string tree-level can induce at one loop level logarithmic corrections of the type eq. (\ref{logdelta}). An explicit computation however is rather difficult to be performed since it requires quantising strings propagating in Calabi-Yau threefolds taking into account the perturbative in $\alpha'$ correction (thus treating it exactly), and it is not within the scope of the present work.

In orbifold compactifications of type IIB orientifolds, the $\alpha'$ correction $\xi$ vanishes. Thus at the leading order in string loop expansion, graviton kinetic terms in eq. (\ref{hatcorrections}) are ten-dimensional and therefore the arguments of \cite{Antoniadis:1998ax} do not apply. However this is not the case at higher orders. The one loop correction $\hat\delta$ receives moduli dependent contributions only from $N=2$ supersymmetric sectors depending on the moduli of the corresponding fixed torus under the action of the orbifold group~{\cite{Antoniadis:1996vw}. For $D7$-branes transverse to the 2-torus $T^2$, it is given by a sum over BPS states corresponding to the open string winding modes where $N=2$ vector multiplets and hypermultiplets contribute with opposite signs. The result depends on the complex structure modulus of the torus but not on its volume and does not contain any logarithmic correction, as expected from our general analysis above. Thus, one loop corrections in the K\"ahler potential cannot lead to logarithmic dependence, in agreement with one-loop results in the literature for orbifolds (see for instance~\cite{Antoniadis:1996vw, Berg:2005ja}). On the contrary, the kinetic function of $D7$ gauge fields which are localised in the transverse dimension receive large corrections that grow logarithmically with the transverse volume, which is calculated explicitly in \cite{Antoniadis:1996vw}. Notice however, that the one loop corrections \cite{Berg:2005ja} contain terms localised on the transverse $T^2$ and thus two loop corrections are expected to diverge logarithmically with its volume, following the argument above. In this case, the correction $\delta$ in eqs.~(\ref{unhated}) and (\ref{logdelta}) should have an additional factor of $g_s^2 \ln u$.

In the following, we will  consider a radiatively corrected K\"ahler potential (\ref{Kahlercorrected}) with eq. (\ref{logdelta}) $\delta=\eta\ln u$, :
\ba 
{\cal K} &=& -\ln(S-\bar S)-2\ln\left( \hat{\cal V}+ {\xi}+ {\eta}\ln u \right)\,.\label{KLogcorrected}
\ea

\subsection{D-terms in the presence of $D7$-branes }

It has been suggested that magnetised branes along $(1,1)$-cycles of the internal compactification space can be used to stabilise the K\"ahler moduli, as an alternative to non-pertubative effects, at a de Sitter vacuum through the induced  D-terms~\cite{Antoniadis:2004pp, Antoniadis:2005nu, Antoniadis:2008uk}. The advantage of magnetic fluxes on $D$-branes, as opposed to non-perturbative effects and $\overline{D3}$-contributions, is that these have an exact string description at weak coupling (i.e. to all orders in $\alpha'$) and can be studied within the standard effective supergravity. In this subsection we will discuss 
the D-term contributions from magnetised $D7$-branes in type IIB superstring theory. We will assume that all complex structure moduli and the axion-dilaton ten-dimensional (10d) field are fixed in a standard way by appropriate 3-form fluxes at a vacuum preserving $N=1$ supersymmetry in four dimensions with weak string coupling. Moreover, we shall consider zero vacuum expectation values (VEVs) for all charged fields and restrict our analysis to the K\"ahler moduli associated with the world- and transverse-volumes of the $D7$-branes; they should all be considered large in string units for the consistency of the effective supergravity description. 

In usual $D7$-brane configurations
representing supersymmetric four-dimensional (4d) effective theories there are stacks of
branes associated with some non-abelian gauge group while it is common that additional 
branes intersect each other. 
A single $D7$-brane spans four compact dimensions  and forms a two-cycle intersection
with any other non-overlapping brane.

We now consider a IIB/F-theory framework with the presence  of intersecting $7$-branes.
Stacks of $D7$-branes are associated with 
gauge groups and we assume a $D$-brane configuration where some anomalous $U(1)$ is present, 
induced by a corresponding magnetic flux.
A 4-cycle K\"ahler modulus $T_a$ associated with the world-volume of the magnetised $D7$-brane
acquires then a charge $Q$ under the $U(1)$ as a shift symmetry along its real component:
$T^a\to T^a+Q\omega$, with $\omega$ the transformation parameter
(the appropriate topological conditions 
for this, are discussed for instance in~\cite{Haack:2006cy}).  In general, we also expect
the existence of complex scalar fields $\phi^J$ carrying charges $Q_J$. 
 
The induced D-term has the generic form dictated by the effective $N=1$ supergravity \cite{Burgess:2003ic, Jockers:2004yj, Achucarro:2006zf, Haack:2006cy}:
\ba
V_D = \frac{g_{D7}^2}{2} \left( iQ\partial_{T^a}{\cal K}(T^a)+\sum_JQ_J\mid\langle \phi^J\rangle \mid^2 \right)^2\label{D7Kahl}
\ea 
where the gauge coupling is fixed by the kinetic function: $\frac{1}{g^2_{D7}}={\rm Im}(T^a)$ and  $\phi^J$ are scalar components
of superfields whose charges $Q_J$ are subject to anomaly cancellation conditions 
(that are automatically satisfied in a consistent string background)~\cite{Antoniadis:2005nu}.
Although in general the VEVs of the scalar fields are on-zero, for our
present purposes we can ignore the matter fields and write (\ref{D7Kahl})
as follows
\ba
V_D=-\frac{d_a}{{2\rm Im}(T^a)} \left(\partial_{T^a}{\cal K}(T^a)\right)^2,
\ea 
in which $d_a = Q^2$~\footnote{See also argument in~\cite{Achucarro:2006zf}.}.

In our convention, we denote the imaginary part of the world-volume K\"ahler modulus $T_a$ as $\tau_a$. The whole 6-dimensional volume can be expressed as sum of triple products of 2-cycle moduli:
\beq \mathcal{V} = \frac{1}{6}\kappa_{abc} v^a v^b v^c, \eeq
where $\kappa_{abc}$ are the triple intersection numbers. In the framework of 3 intersecting $D7$-branes, we take 2-cycle $v^a$ as the  transverse volume modulus of each $D7$-brane with world-volume $\tau_a$:
\beq v_a = \frac{\cal V}{\tau_a}, \eeq
 and take $\kappa_{abc}$ as $\epsilon_{abc}$ for simplicity. Then the volume can be expressed as
 \beq \mathcal{V} = v_1 v_2 v_3 = \sqrt{\tau_1\tau_2\tau_3}\eeq

\section{ Volume Stabilisation from Intersecting $D7$-Branes}

In this section we investigate the implications of $D7$-branes on the stabilisation of K\"ahler moduli. Starting with the simplest case, we introduce
only one $D7$-brane and observe that this is not adequate to stabilise all moduli. Then, we proceed with the inclusion of three intersecting $D7$-branes. 

\subsection{A Single $D7$-Brane}
We start with a single space-time filling $D7$-brane and assume that all the complex structure moduli and the axion-dilaton are stabilised by fluxes. The K\"{a}hler modulus can be divided into the world volume part of the $D7$-brane $\tau$ and the transverse part $u$. Both $\tau$ and u are real 4-cycle volumes. Then we can write the compactifacation volume $\mathcal{V}$ in terms of the two K\"{a}hler  moduli as follows:
\beq \mathcal{V} = \tau \sqrt{u}. \eeq
The no-scale structure is broken by perturbative corrections: $\alpha^\prime$ world-sheet corrections and string loop corrections. The K\"{a}hler potential now takes the general form:
\beq {\cal K} = -2\, \ln (\tau \sqrt{u}+ \xi + \eta \textrm{ln}(u)),\eeq
The string loop correction term $\eta \ln(u)$ is of course valid in the perturbative region:
\beq |\eta \textrm{ln}(u)| < \tau \sqrt{u}\eeq
The corresponding F-term potential with a superpotential $\mathcal{W}_0$ is:
\beq V_F =  \frac{\mathcal{W}_0^2 (-8\eta + 3\xi +3\eta \textrm{ln}(u))} {(8\eta + 2\tau \sqrt{u} - \xi - \eta \textrm{ln}(u)) (\tau \sqrt{u}+ \xi + \eta \textrm{ln}(u))^2}. \eeq

In the large volume expansion, we can compute the derivative with respect to $u$:
\beq \frac{d V_F (\tau, u)}{d u} = -\eta \mathcal{W}_0^2 \frac{3 (-10 + 3 \textrm{ln}(u))}{4\tau^3 u^{5/2}} + O(\eta^2) + O(\xi). \eeq
We find that for $\eta$ being negative, the potential has a minimum in the $u$ direction. Thus, the string loop correction $\eta \textrm{ln}(u)$ can stabilise the transverse direction of the $D7$-brane. However, for the volume part $\tau$, the first derivative doesn't show the stabilisation. Indeed, in the appendix we show that even in the presence of an uplifting D-term, there is no dS minimum with just perturbative corrections for a single $D7$-brane. 

\subsection{Stabilisation of the total volume by three intersecting $D7$-branes}
In the same way, we can get the F-term potential for 2 non-parallel $D7$-branes and find that there is always one K\"{a}hler modulus which is not stabilised. Thus, in order to stabilise all the K\"{a}hler moduli, we should consider that there exist at least three non-parallel (magnetised) $D7$-branes. This corresponds to 3 intersecting $D7$-branes which is quite general in string model building. In the following, we neglect the $\alpha^\prime$ correction and consider only the string loop correction. The general K\"{a}hler potential can be written as:
\bea
 {\cal K} &&= -2 \textrm{ln}\, (\sqrt{\tau_1 \tau_2 \tau_3} +  \sum_i 2\eta_i \textrm{ln}(\frac{\mathcal{V}}{\tau_i}))\\
&&= -2 \textrm{ln}\, (\sqrt{\tau_1 \tau_2 \tau_3} +  \sum_i \eta_i^\prime \textrm{ln}(\tau_i)), \qquad \eta_a^\prime = \sum_i\eta_i - 2\eta_a.\eea
Each $\tau_i$ corresponds to the world volume of one $D7$-brane real 4-cycle. We calculate the first derivative with respect to either $\tau_a$:
\beq \frac{d V_F (\tau_1, \tau_2, \tau_3)}{d \tau_a} = \mathcal{W}_0^2 \frac{3 (\sum_{i\neq a} (8\eta^\prime_i - 3\eta_i^\prime \textrm{ln}(\tau_i)) + 10\eta^\prime_a - 3 \eta^\prime_a \textrm{ln}(\tau_a))}{4\prod_{i\neq a}\tau_i^{\frac{3}{2}} \tau_a^{\frac{5}{2}} } + O(\eta^{\prime2}). \label{Fterm1stderivative}\eeq

The minimisation condition from the three directions in eq. (\ref{Fterm1stderivative}) then shows that a minimum only exists for the total volume $\mathcal{V}$ if
\beq \eta_1 = \eta_2 = \eta_3 = \eta_\tau < 0. \label{universaleta}\eeq
The other two directions, which can be thought of as the ratios between $\tau_1$, $\tau_2$ and $\tau_3$, remain flat, since under the condition (\ref{universaleta}), the K\"{a}hler potential and the corresponding F-term potential  only depend on $\mathcal{V}$. Indeed, the K\"ahler potential is:
\beq 
{\cal K} = -2 \textrm{ln} (\mathcal{V} + 2\eta_\tau \textrm{ln} (\mathcal{V}))~,\label{Kpotentialuniversal}
\eeq
and  the F-part of the effective potential is
\bea
V_F (\mathcal{V}) &&= -\frac{3\eta_\tau \mathcal{W}_0^2 (2\eta_\tau + 4\mathcal{V} + 4\eta_\tau \textrm{ln}(\mathcal{V})  - \mathcal{V} \textrm{ln}(\mathcal{V} ))}{(\mathcal{V}  + 2\eta_\tau\textrm{ln}(\mathcal{V}))^2 (6\eta_\tau^2 + 8\eta_\tau\mathcal{V} + \mathcal{V}^2 + \eta_\tau (4\eta_\tau - \mathcal{V}) \textrm{ln}(\mathcal{V} ))} \label{Ftermpotential}\\
&&= \frac{\eta_\tau \mathcal{W}_0^2}{\mathcal{V}^3} (3 \textrm{ln} (\mathcal{V}) - 12) + O(\eta_\tau^2)\label{Ftermpotentiallimit}
\eea
\begin{figure}[H]
	\centering
	\includegraphics[width=0.80\columnwidth]{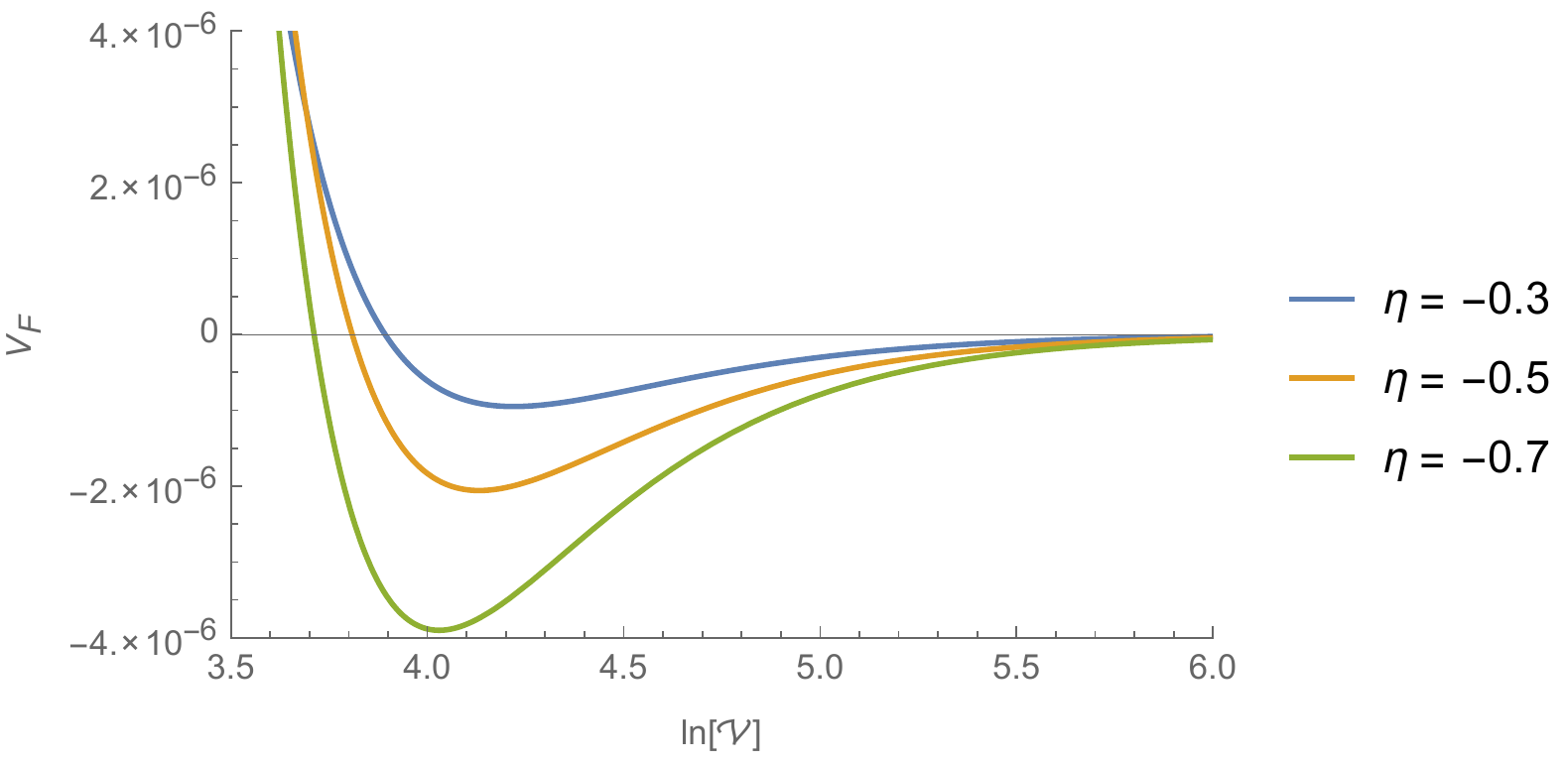}
	\caption{
		\footnotesize
		{The scalar potential of eq. (\ref{Ftermpotential}) for the choice  $\mathcal{W}_0$ = 1.}
	}
	\label{1variableFtermpotential}
\end{figure}
The first derivative of eq. (\ref{Ftermpotential}) shows that the minimum is independent of the $\eta_\tau$ parameter in the large volume limit:
\beq
\frac{d V_F (\mathcal{V})}{d \mathcal{V}} = -\eta_\tau\frac{3 \mathcal{W}_0^2 (3\textrm{ln} (\mathcal{V}) - 13)}{\mathcal{V}^4} + O (\eta_\tau^2). 
\eeq
The potential~(\ref{Ftermpotential}) is plotted in  Fig.~\ref{1variableFtermpotential}.\\
At this minimum, the supersymmetric condition is not satisfied:
\beq D_\mathcal{V} \mathcal{W}_0\mid_{\mathcal{V}_{min}} = \partial_\mathcal{V} K\mid_{\mathcal{V}_{min}} \mathcal{W}_0 = -\frac{2\mathcal{W}_0}{\mathcal{V}_{min}} + O(\eta_\tau) \neq 0,\eeq
so supersymmetry is spontaneously broken. The minimum is not stable due to the two undetermined directions and a tiny deviation from condition~(\ref{universaleta}) would destabilise the total volume. We will discuss the stabilisation of the ratios in the next section where the condition~(\ref{universaleta}) is not necessary. 

There is a similar form to eq. (\ref{Kpotentialuniversal}) in the classical large volume scenario. The equation of motion of the small cycle $\tau_s$ from eq.~(\ref{LVS}) leads to
\beq \tau_s \propto \textrm{ln}(\mathcal{V}) \propto \textrm{ln}(\tau_b).\eeq

\subsection{D-term uplifting and ratios stabilisation}
In order to stabilise the ratios and find a dS vacuum, we introduce  D-terms  emerging from a magnetic flux on each $D7$-brane. These  depend  on the corresponding world volume modulus $\tau_a$:
\bea V_{D_a} &=& \frac{d_a}{\tau_a} \left(\frac{\partial K}{\partial {\tau_a}}\right)^2 \,=\, d_a\frac{(\mathcal{V} + 2\eta^\prime_a)^2}{\tau_a^3 (\mathcal{V} + \sum_i\eta^\prime_i \textrm{ln}(\tau_i) )^2}\label{VD-term}\\ &=& \frac{d_a}{\tau_a^3} + O(\eta_i). \label{VD-termlimit}\eea
For  simplicity, we still use the condition (\ref{universaleta}) to calculate the minimum. Notice that it is not a necessary condition once D-terms are included from all three $D7$-brane stacks, leading to a global minimum for all the K\"{a}hler moduli. 

We choose $\tau_1, \tau_2$ and $\mathcal{V}$ as the 3 independent dynamical variables. The sum of the F-term potential (\ref{Ftermpotentiallimit}) and D-term potentials (\ref{VD-termlimit}) in the large volume limit becomes:
\beq V_{sum} = \frac{\eta_\tau \mathcal{W}_0^2}{\mathcal{V}^3} \left(3 \textrm{ln} (\mathcal{V}) - 12\right) + \frac{d_1}{\tau_1^3} + \frac{d_2}{\tau_2^3} + \frac{d_3 \tau_1^3 \tau_2^3}{\mathcal{V}^6}. \eeq
The minimisation conditions of $\tau_1$ and $\tau_2$ lead directly to
\bea 
\tau_1^3 &=& \left(\frac{d_1^2}{d_2 d_3}\right)^{\frac{1}{3}} \mathcal{V}^2 \nnmb\\
\tau_2^3 &=& \left(\frac{d_2^2}{d_1 d_3}\right)^{\frac{1}{3}} \mathcal{V}^2.\label{minimizationoftau1tau2}
\eea
Substituting these expresssions into the minimisation condition of $\mathcal{V}$ we get:
\beq \eta_\tau \mathcal{W}_0^2 (13 - 3 \textrm{ln} (\mathcal{V})) = 2(d_1d_2d_3)^\frac{1}{3} \mathcal{V}. \label{minimizationofV}\eeq
There are two conditions that must be satisfied  in order to get a dS minimum. 

\begin{itemize}
\item
The first is that  there should exist two real solutions of eq. (\ref{minimizationofV}) where the smaller one corresponds to a minimum and the larger one corresponds to a maximum. Indeed, by doing a change of variables in eq. (\ref{minimizationofV}), we get:
\beq z e^{z} = \frac{2 e^{\frac{13}{3}} (d_1d_2d_3)^{\frac{1}{3}}} {3\eta_\tau \mathcal{W}_0^2}\quad;\quad z = \frac{13}{3} - \textrm{ln} (\mathcal{V}).\label{Vtoz}\eeq
The first equation above has two solutions for $z$ negative while the function $ze^z$ has a minimum at $z=-1$. Thus, the right hand side should be between $-e^{-1}$ and 0 that requires:
\beq -\frac{3 e^{-\frac{16}{3}}}{2} \simeq - 0.007242 < \frac{(d_1d_2d_3)^{\frac{1}{3}}}{\eta_\tau \mathcal{W}_0^2} < 0.\label{2extremecondition}\eeq
The smaller solution $\mathcal{V}_0$ of eq. (\ref{minimizationofV}) is:
\beq \mathcal{V}_0 = e^{\frac{13}{3} - W \left[\frac{2 e^{\frac{13}{3}} (d_1d_2d_3)^{\frac{1}{3}}} {3\eta_\tau \mathcal{W}_0^2}\right]}\, ,\eeq
in which W is the  Lambert W-Function. 

\item
The second condition is that the potential should be positive at the minimum. Using eqs.~(\ref{minimizationoftau1tau2}) and (\ref{minimizationofV}), we can express the potential at the minimum in a simple form:
\beq V_{sum}^{min} = \frac{\eta_\tau \mathcal{W}_0^2}{\mathcal{V}_{0}^3} + \frac{(d_1d_2d_3)^{\frac{1}{3}}}{\mathcal{V}_0^2} > 0\eeq
Solving it numerically gives a new constraint
\beq \frac{(d_1d_2d_3)^{\frac{1}{3}}}{\eta_\tau \mathcal{W}_0^2} < -0.006738,\eeq
which is consistent with the inequalities (\ref{2extremecondition}) and together lead to
\beq
 -0.007242 < \frac{(d_1d_2d_3)^{\frac{1}{3}}}{\eta_\tau \mathcal{W}_0^2} < -0.006738.
\eeq
Within this range, we can get approximately the order of magnitude of the volume $\mathcal{V}_0$:
\be
 \textrm{ln} (\mathcal{V}_0) \simeq 5, 
 \ee
which corresponds to a Grand Unification Theory (GUT) scale compactification volume.
\end{itemize}

\noindent 
We show an example in Fig.~\ref{Vsum2parameters}, in which we take $d_1 = d_2$, thus $\tau_1 = \tau_2$ at minimum according to eq. (\ref{minimizationoftau1tau2}).

\begin{figure}[H]
	\centering
	\includegraphics[width=0.60\columnwidth]{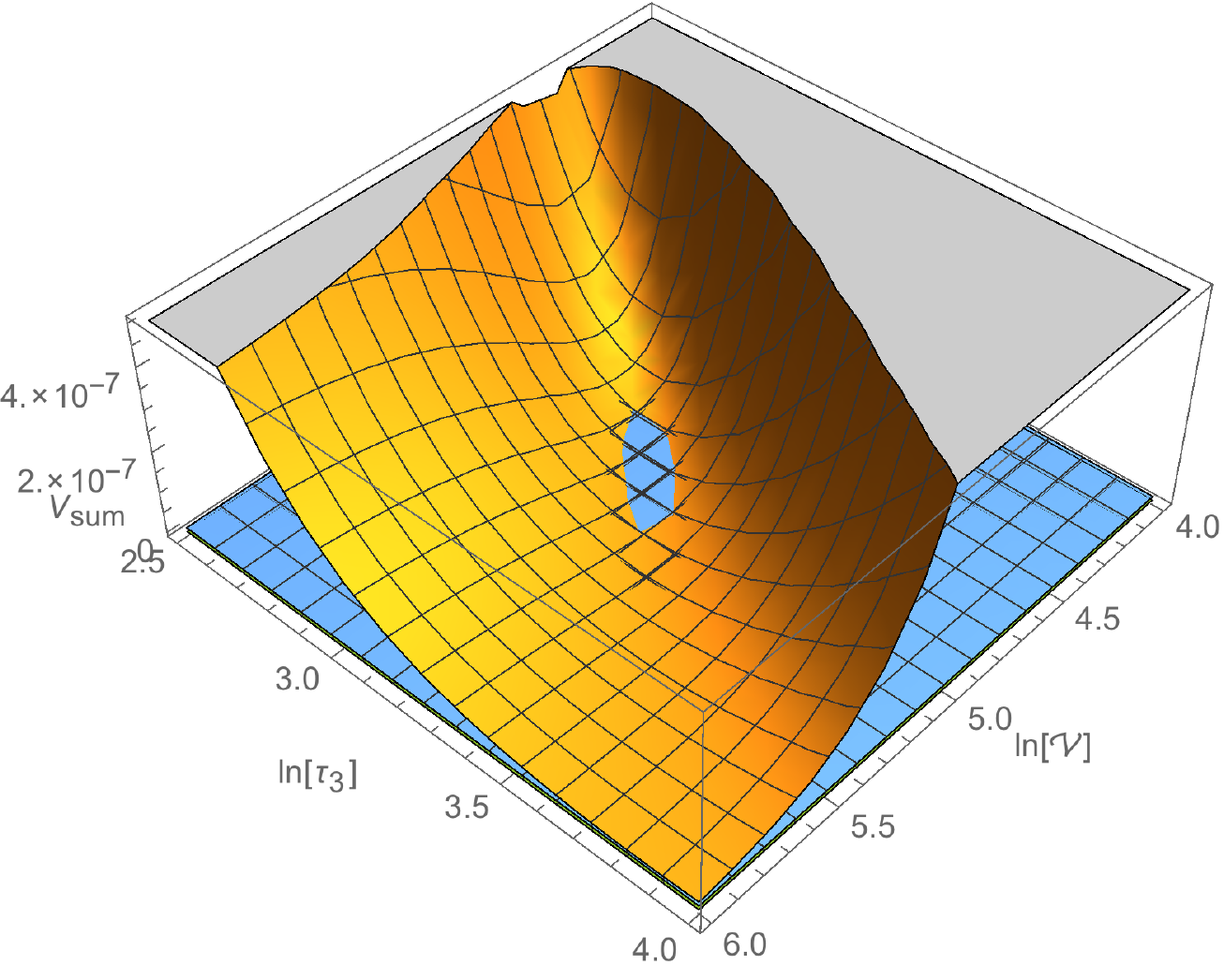}
	\caption{
		\footnotesize
		{Sum of the F- and D-term contributions to the potential (\ref{Ftermpotential}) and (\ref{VD-term}) in terms of $\textrm{ln}({\mathcal{V}})$ and $\textrm{ln} (\tau_3)$ with the choice $\mathcal{W}_0$ = 1, $\eta_\tau = -0.4,\, d_1 = d_2 = 0.00375$, $d_3 = 0.0018$. The blue area corresponds to $V = 10^{-8}$ plane.}
	}
	\label{Vsum2parameters}
\end{figure}

\section{Conclusions}

Moduli stabilisation in string theory is a long standing issue and despite the 
significant progress that has been made during the last two decades, the proposed solutions are still far from being  conclusive. The main ingredients of the existing scenarios are backgound fluxes, string loop-corrections and  non-perturbative effects. 

The key point towards a convincing solution is to implement a realistic  dynamical mechanism  which generates a scalar potential and  provides masses to the various massless scalar fields emerging in string compactifications.
In this  work, we have studied this problem in the framework of IIB/F-theory  compactifications and we have proposed a new geometric  mechanism which dispences with the use of non-perturbative effects.  We have  considered  configurations,  where the main ingedients are  intersecting $D7$-branes equipped with internal magnetic  fluxes
which have an exact description to all orders in $\alpha'$, and   we have investigated  their implications on the  stabilisation of the K\"ahler moduli. 
More concretely,  assuming that the VEVs of the complex structure moduli and of the axion-dilaton field are already fixed by supersymmetry conditions, we examined  the modifications of the K\"ahler potential arising from perturbative $\alpha' $ and loop corrections.   Elaborating on the essential  features of $D7$-branes in the configuration of the compact space,  we concluded that in the transverse large volume limit of dimension two,  the effective action receives loop corrections which  are logarithmically divergent. 
In effect, the Calabi-Yau volume in the K\"ahler potential receives corrections  which  display  logarithmic dependence on the size of the transverse to the $D7$-branes directions. This is in contrast to the $\alpha'$ correction which induces just  a shift to the volume by a constant parameter $\xi$.

In addition,  magnetised $D7$-branes, have  significant implications on the stabilisation of the K\"ahler moduli and, at the same time, they can naturally ensure the existence of a dS minimum. More precisely, 
magentised $D7$-branes are associated with anomalous $U(1)$ symmetries which are also a source of  D-terms, that depend on the world-volume of the corresponding $D7$-branes. These contributions to the effective potential can stabilise the ratios between each world-volume modulus and the total volume and, thus, they work as an uplift mechanism to realise de-Sitter minima.  To show this we have computed the scalar potential and performed a detailed analysis, where we found that in the case of one $D7$-brane the logarithmic	shift of the volume stabilises only the modulus of the transverse space. 
Stabilisation of the total volume is achieved only in the presence of at least 
three intersecting $D7$-branes, which span all six dimensions of the compact space. Interestingly, in this scenario, non-perturbative corrections are not necessary.

The realisation of this geometric stabilisation mechanism and the uplifting is a viable scenario in  F-theory~\cite{Vafa:1996xn}  
where intersecting 7-branes are a natural phenomenon which also has additional attractive features  beyond the present context. For instance, one could use the F-theory model building \cite{Beasley:2008dc} to realise the Standard Model.

\section*{Acknowledgements}
This work was supported in part by the Swiss National Science Foundation, in part by Labex ``Institut Lagrange de Paris'' and in part by a CNRS PICS grant. G.K.L. would like to thank the  LPTHE in Paris and the ITP in Bern
for their kind hospitality while Y.C. would like to thank ITP, where part of the work was completed. Y.C. also thanks Karim Benakli, Emilian Dudas and Mark D. Goodsell for discussions.

\section*{Appendix}
The D-term potential from a $U(1)$ magnetic flux on a single $D7$-brane has the form:
\beq V_D = d\frac{u}{\tau (\tau \sqrt{u}+ \xi + \eta \textrm{ln}(u))^2}.\label{Eq:VD}\eeq
The corresponding scalar potential is:
\beq V = \frac{\mathcal{W}_0^2 (-8\eta + 3\xi +3\eta \textrm{ln}(u))} {(8\eta + 2\tau \sqrt{u} - \xi - \eta \textrm{ln}(u)) (\tau \sqrt{u}+ \xi + \eta \textrm{ln}(u))^2} + d\frac{u}{\tau (\tau \sqrt{u}+ \xi + \eta \textrm{ln}(u))^2}.\eeq
In order that the scalar potential has a dS minimum, one necessary condition is that for the direction along $\mathcal{V} = \tau\sqrt{u}$, there exists a dS minimum when $u$ is a constant. It is the same for the direction along $\tau$ since they only differ by a factor of $\sqrt{u}$ which is a positive constant. We can thus write the potential in terms of $\mathcal{V}$ and take the other parameters including $u$ as constants:
\bea V(\mathcal{V}) &=& \frac{a}{(\mathcal{V} - b) (\mathcal{V} - c)^2} + \frac{e}{\mathcal{V}(\mathcal{V} - c)^2}\\
&=& \frac{(a+e)\mathcal{V} - b e} {\mathcal{V} (\mathcal{V}-b) (\mathcal{V}-c)^2} \eea

The potential should be positive when $\mathcal{V} \rightarrow + \infty$. So $(a + e)$ should be positive. Since we only consider the existence of a dS minimum, independently of the overall normalisation of the potential, we can divide it by $(a + e)$ and define a new parameter $f = \frac{b e}{a + e}$. The potential now becomes:
\bea
 \frac{ V(\mathcal{V})}{a + e} = \frac{\mathcal{V} - f} {\mathcal{V} (\mathcal{V}-b) (\mathcal{V}-c)^2} 
 \eea
and has three singularities at $0, b$ and $c$. The dS minimum should lie in the branch outside these three singularities. Thus, we define a new parameter 
\beq g = max (0, b, c),\eeq and only consider the region:
\beq\mathcal{V} > g.\eeq
First, consider the case that $f > g$; we  find that the potential becomes negative in the range $g < \mathcal{V} < f$, which means there is no dS minimum. We turn to the case $f \leq g$ and calculate the first derivative with respect to $\mathcal{V}$:
\beq \frac{dV (\mathcal{V})}{d\mathcal{V}} /(a + e) = \frac{1} {\mathcal{V} (\mathcal{V}-b) (\mathcal{V}-c)^2} -  \frac{\mathcal{V} - f} {\mathcal{V}^2 (\mathcal{V}-b) (\mathcal{V}-c)^2} -  \frac{\mathcal{V} - f} {\mathcal{V} (\mathcal{V}-b)^2 (\mathcal{V}-c)^2} - \frac{2(\mathcal{V} - f)} {\mathcal{V} (\mathcal{V}-b) (\mathcal{V}-c)^3}.\eeq
Suppose $b$ is the largest singularity $b = g$. The first derivative becomes:
\beq \frac{dV (\mathcal{V})}{d\mathcal{V}} /(a + e) = \frac{f - b} {\mathcal{V} (\mathcal{V}-b)^2 (\mathcal{V}-c)^2} -  \frac{\mathcal{V} - f} {\mathcal{V}^2 (\mathcal{V}-b) (\mathcal{V}-c)^2} - \frac{2(\mathcal{V} - f)} {\mathcal{V} (\mathcal{V}-b) (\mathcal{V}-c)^3}.\eeq
Note that all the terms above are negative in the region $f\leq g < \mathcal{V}$. Thus, there is no minimum in this region. The same results hold for $0$ or $c$ to be the largest singularity $g$. Thus, no dS minimum exists in the physical region of the parameter space.

\end{document}